# Authenticating Transactions using Bank – Verified Biometrics

Sankalp Bagaria
sankalp[dot]nitt[at]gmail[dot]com

**Abstract** - In this paper, we propose a scheme by which banks can collect and verify biometric data eg, fingerprints, directly from their customers and use it to authenticate their transactions made through PoS/ ATM/ online console. We propose building a network of computers called BioNet to allow such transactions to be made online across the world. A BioNet server will be able to do 4 million transactions per second using GPU.

***I. Introduction*** Banks do a stringent check on their customers. If they are allowed to collect finger-prints of their customers, a secure and convenient alternative to debit cards/ credit cards/ online passwords etc can be reached. Biometrics eg. fingerprint matching, is a more secure technology than that of credit card, debit card and passwords. We propose that BioNet be built on the lines of Mastercard, Visa card etc to achieve this. This will also remove the need to remember passwords or carry cards as transactions would be made by swiping the fingers on PoS. The risks of card-threat and card-fraud would also be removed.

The remainder of the paper is organized as follows. Section II talks about the role of biometrics in banks. Section III claims fingerprint is sufficient for authenticating transactions. Section IV introduces BioNet as an alternative to Visa card, Mastercard etc. and Section V discusses BioNet Server details. The article concludes in Section VI and discusses future work in Section VII.

***II. Bank and Biometrics*** The banks do a stringent check, when a customer opens bank accounts or applies for a credit card. So, if finger-prints are collected by banks when a customer is opening an account or applying for a credit card etc, it would result in a reliable and useful database. Banks already have a lot of sensitive data like our signatures, account details, financial transactions etc which they protect. This will help us in finding a more secure and convenient alternative to Visa card/ Master card. This will help us prevent losses due to card-theft, card-fraud across the world.

***III. Fingerprint is sufficient.*** We also claim that the finger-print is sufficient to authenticate transaction at point-of-sale terminal in shops/ ATMs. Fingerprints are a more convenient and secure way of authenticating transactions than credit card/ debit card. The finger-print details would be transmitted from one point to another in a secure way throughout the transaction.

Customer would enter his/ her 4-digit pin number and swipe his/her finger at PoS. PoS will encrypt the fingerprint and send it to the merchant's bank. The merchant's bank will decrypt it, read the pin number and send the finger-print details to the appropriate BioNet server (built like Mastercard [5]/ Visa card [6] network) after encrypting it. The BioNet server will decrypt the finger-print details and match the customer's finger-print with the finger-print samples stored in it. If there is a match, it will read the customer's details stored in its memory and locate the customer's bank and branch. Customer's bank will check account details and allow/ deny transaction. This verdict will be returned to merchant's bank via BioNet. Merchant's bank will advise PoS to allow/ deny transaction. The whole transaction would be done in a secure way by using symmetric key encryption over the links.

Not only shops, biometric sensors can be installed on desktops [3] and mobile phones for online transactions through Internet. It would prevent someone seeing the card details over the shoulder of the customer wen he/she enters the card details in the terminal. It will also prevent a keylogger from intercepting typed card details.

Another advantage is that police may contact banks/ BioNet to red-flag a convict's account and that may be used to locate convict when he/she swipes his/her finger at any PoS. One more advantage of this scheme is that there will be no need to carry credit card/debit card. Card-theft and card-fraud will also be avoided. It can also be used in rural areas also where literacy level is low.

*VI. The BioNet* The BioNet will be a network of servers like that of Mastercard/ Visa card. Each customer of participating banks and financial organizations will be assigned a 4-digit pin number which he/she will enter at PoS/ ATM/ mobile/ computer terminal. This number will identify the BioNet server to which his/her fingerprint sample has been sent by his/ her bank. When the customer enters his/ her pin number and swipes his finger, the encrypted finger-print detail will be sent to appropriate BioNet server securely. There are 10,000 servers possible in the BioNet for the 4-digit pin number. Each of these 10,000 servers will have fingerprinting algorithm running on it. Minutia Cylinder code of BioLab [13] uses GPU to achieve 4 million matches/sec. If a customer at the PoS has many records in the server's database, he/she may be asked to choose the particular bank account from which the amount needs to be debited from.

*VII. The BioNet Server* Each BioNet server, identified by its pin number, will run fingerprint matching algorithm. It will receive the fingerprint of the customer from the merchant's bank, run the fingerprint algorithm and check if it matches one of the fingerprint samples stored in it. Banks would have stored the finger-prints in the respective servers before-hand. If there is a match, it will read the details of customer's bank and branch from its database. Then it will create connection with the appropriate bank and branch for further processing of the transaction. If there is no match, it will advise the PoS to deny the transaction.

BioLab has used GPU to achieve 4 million fingerprint matches/ second using Cuda. This means each of the BioNet's 10000 servers can match 4 million fingerprints/ second. If each of these 10,000 servers had 1 million finger-print samples stored to match, it would mean 10 billion fingerprints total (a number larger than the world population). Each server will then be able to match 4 fingerprints in one second in the worst case. This will map to about 40,000 transactions per second in the worst case.

This could be further improved by having a cluster of servers being assigned same pin number such that a subset of 1 million fingerprints is assigned to each server in the cluster. The effective throughput of the cluster of servers will increase this way.

*VIII. Conclusion* In this article, we discussed how we would have a more secure and convenient alternative to debit card/ credit card/ online passwords across the world. We proposed that the banks be allowed to collect biometric data of their customers and use that to authenticate transactions – be it through PoS, ATM or online console. This will prevent losses due to card-theft, card-fraud, hacking online passwords etc.

We presented BioNet in this article for doing what Visa card and Master card do now for credit card and debit cards. Such a scheme, based on finger-print matching, would be more secure and convenient to use than card-technology. BioNet will be a network of servers which would be running finger-print matching algorithm. Multi – core CPUs working with GPUs can give the required throughput using the present – day technology. We showed that a network of around 10,000 servers can service to the whole world population.

*IX. Future Work* BioNet has to be built. Finger-print matching is highly parallelizable. Faster algorithms can be written that can make use of parallel architecture. Later on dedicated chips can be built to obtain optimum throughput. A hierarchy architecture like decision tree can further optimize finger-print matching. Bio – hashing has to be further researched.